\documentstyle[epsfig]{elsart}

\begin{document}
\begin{frontmatter}

\title{ Thin Ice Target for $^{16}$O$(p,p')$ experiment
}

\author[Kyoto]{T.~Kawabata\thanksref{EMAIL}},
\author[Konan]{H.~Akimune}, 
\author[RCNP]{H.~Fujimura}, 
\author[Osaka]{H.~Fujita},
\author[Osaka]{Y.~Fujita},
\author[RCNP,JAERI]{M.~Fujiwara}, 
\author[RCNP]{K.~Hara}, 
\author[RCNP]{K.~Hatanaka},
\author[Himeji]{K.~Hosono},
\author[Kyoto]{T.~Ishikawa},
\author[Kyoto]{M.~Itoh}, 
\author[RCNP]{J.~Kamiya},
\author[Kyoto]{M.~Nakamura}, 
\author[RCNP]{T.~Noro}, 
\author[RCNP]{E.~Obayashi}, 
\author[Kyoto]{H.~Sakaguchi},
\author[Osaka]{Y.~Shimbara}, 
\author[Kyoto]{H.~Takeda}, 
\author[Kyoto]{T.~Taki\thanksref{taki}}, 
\author[Tokyo]{A.~Tamii}, 
\author[SPring8]{H.~Toyokawa},
\author[Kyoto]{N.~Tsukahara}, 
\author[Kyoto]{M.~Uchida}, 
\author[Osaka]{H.~Ueno\thanksref{ueno}}, 
\author[RCNP]{T.~Wakasa}, 
\author[Konan]{K.~Yamasaki}, 
\author[Kyoto]{Y.~Yasuda}, 
\author[RCNP]{H.P.~Yoshida}, and
\author[RCNP]{M.~Yosoi}

\address[Kyoto]{Department of Physics, Kyoto University, 
Kyoto 606-8502, Japan
}
\address[Konan]{Department of Physics, Konan University,
Kobe, Hyogo 658-8501, Japan
}
\address[RCNP]{Research Center for Nuclear Physics, 
Osaka University, Ibaraki, Osaka 567-0047, Japan
}
\address[Osaka]{Department of Physics, Osaka University,
Toyonaka, Osaka 560-0043, Japan
}
\address[JAERI]{Advanced Science Research Center,
Japan Atomic Energy Research Institute,
Tokai-mura, Ibaraki 319-1195, Japan
}
\address[Himeji]{Department of Engineering, Himeji Institute of Technology,
Hyogo 678-1297, Japan
}
\address[Tokyo]{Department of Physics, University of Tokyo, Hongo,
Tokyo 113-0033, Japan
}
\address[SPring8]{Japan Synchrotron Radiation Research Institute, 
Hyogo 679-5198, Japan
}
\thanks[EMAIL]{
E-mail: kawabata@ne.scphys.kyoto-u.ac.jp
}
\thanks[taki]{
Present address: Asaka Technology Development Center, 
Fuji Photo Film Co., Ltd., Asaka, Saitama 351-8585, Japan}
\thanks[ueno]{
Present address: Insitute for Chemical and Physical Research (RIKEN),
Wako, Saitama 351-0198, Japan}

\begin{abstract}
A windowless and self-supporting ice target is described. An ice sheet
with a thickness of 29.7 mg/cm$^2$ cooled by liquid nitrogen was placed at
the target position of a magnetic spectrometer and worked stably in the
$^{16}$O$(p,p')$ experiment at $E_{p}=392$ MeV.  Background-free
spectra were obtained.

\begin{flushleft}
\end{flushleft}
\end{abstract}
\end{frontmatter}


\section{Introduction}

The  $^{16}$O nucleus is an important target for the
study of nuclear physics because of the spin-saturated closed-shell
property. Extensive measurements of $(e,e')$ \cite{voegler}, 
$(p,p')$ \cite{larson,seifert,djalali}, $(p,n)$ \cite{mercer}, $(p,d)$
\cite{abegg}, and $(p,2p)$ \cite{miller}
reactions have been devoted to investigate various nuclear properties,
{\it e.g.} excited resonance structures, single particle or hole states
and, effective interactions in the nuclear medium.

It is, however, not easy to prepare an $^{16}$O target due to its
gaseous property. Table ~\ref{table:targets} summarizes the types of
the oxygen targets and their characteristics.   
Metal oxides are commonly used as oxygen targets since their preparation in 
a chemical process is relatively easy.  The solid compositions of BeO 
and Li$_2$O are conveniently used in many experiments.
Recently, a thin SiO$_2$ (glass) target has been successfully employed 
for the coincidence measurement of decay particles from the deep-hole
state excited by the $(p,2p)$ reaction \cite{yosoi}. In order to extract
the oxygen events, however, the events from the metal base should be
measured separately and subtracted.  
This procedure often includes intrinsic problems concerning the
statistics and quality of spectra. 
When we use a thin target, the evaluation of the contamination in the
metal base is also difficult.  

Oxygen gas kept in a cell is also used as a target.  
At backward scattering angles, the events from the cell  
windows can be removed by means of a double-slit collimator.  
At forward angles, however, this is not so easy; again the subtraction 
procedure is needed.  
Moreover,
the effective thickness changes with scattering angles depending on the
slit geometry.   The long cell length needed to assure enough
target thickness makes it difficult to use the ray-tracing technique
in combination with a magnetic spectrometer.  
A ``water fall'' target, in which water flows membranously
in a target cell, has been used at several facilities \cite{water1,water2}.
This target is normally thick, more than 25 mg/cm$^2$.  

A windowless gas target can be obtained by injecting a supersonic gas
jet into a vacuum chamber and immediately evacuating the gas by pumps
\cite{gasjet1,gasjet3,gasjet6}, but this gas-jet target is too thin
(maximum $\sim\,$1 mg/cm$^2$). In addition, a gas-jet target is
very expensive since it requires a large pumping and gas recycling
system due to a high gas-flow rate ($\sim25$ m$^3$/h). A differential 
pumped target, which keeps vacuum with orifices by reducing gas flow,
is also windowless \cite{diffpump}. However, it is thinner than a
gas-jet target. A differential pumped target is usually 
used in low energy experiments. 

Recently new types of experiments have become feasible to measure the decay
of charged particles from highly excited states in coincidence with
$(p,p')$, $(p,2p)$ and ($^3$He,$t$) events at intermediate 
energies~\cite{yosoi,aki94,aki95,aki00,toyokawa}.
Information from these measurements may be combined with those from 
polarization transfer observables \cite{tam99} 
in order to study microscopic structures of highly excited states.
The detection of decay charged particles requires 
a thin target to minimize the energy loss. 
In addition, some free space is needed around the target to 
install detectors for charged decay particles. 

An frozen H$_2$O target 
was developed for the use as an oxygen target
at Los Alamos National Laboratory (LANL) \cite{ice1,ice2}, and   
it was successfully used in a $(p,n)$ experiment~\cite{mercer}. 
Unfortunately, since the LANL target was rather thick
($\sim300$~mg/cm$^2$) and included window foil to prevent the ice from
sublimation, it is not suitable for the detection of charged decay
particles.  Frozen H$_2$O targets are, however, attractive because of the high 
density and also because 
hydrogen events can be removed easily by using a large
difference of the kinematical effects between hydrogen and oxygen. 
In addition, there is no excited state of hydrogen in our excitation
energy region ($E_x {\leq} {\sim}30$ MeV). 
It is estimated that sublimation loss of ice in the vacuum can be
neglected if the ice sheet-target is kept under very low temperature.  
Vapor-pressure of H$_2$O is 
of the order of $10^{-5}$ Pa at 140 K and decreases exponentially at
lower temperature. Window-foil, therefore, is not needed below 140 K. 

Possible heating up of the ice target by the irradiation of the beam is
estimated.
The temperature distribution $T(\vec{r})$ on the ice target is
described by the following equation at the stationary state;
\begin{equation}
\label{eq:thermal}
- {\rm div}(\kappa\nabla T(\vec{r})) = q(\vec{r}), 
\end{equation}
where $\kappa$ denotes the thermal conductivity and $q(\vec{r})$
denotes the volumetric heat generation rate. 
The rise of the temperature by the energy loss of the proton beam 
was estimated under the simple assumptions.  
If the axial symmetry of the target disk is assumed,
Eq.~(\ref{eq:thermal}) can be easily solved with the boundary
condition that $T=90$ K at the edge of the copper frame and
$r\frac{dT}{dr}=0\,$K at the center of the target.
Temperatures at the center of the 30 mg/cm$^2$-thick ice target
obtained from Eq.~(\ref{eq:thermal}) are 93.4 K and 97.8 K
for the proton beam intensities of 10 nA and 100 nA, respectively. 
The calculated temperatures are much lower than the critical 
temperature of 140 K.

In this article, we describe a windowless and self-supporting 
ice target developed at the Research Center for Nuclear Physics
(RCNP), Osaka university. The performance of the ice target as the
oxygen target  examined in a $(p,p')$ experiment is presented.

\section{Preparation of Ice Target}

\begin{figure}
\begin{center}
\epsfig{file=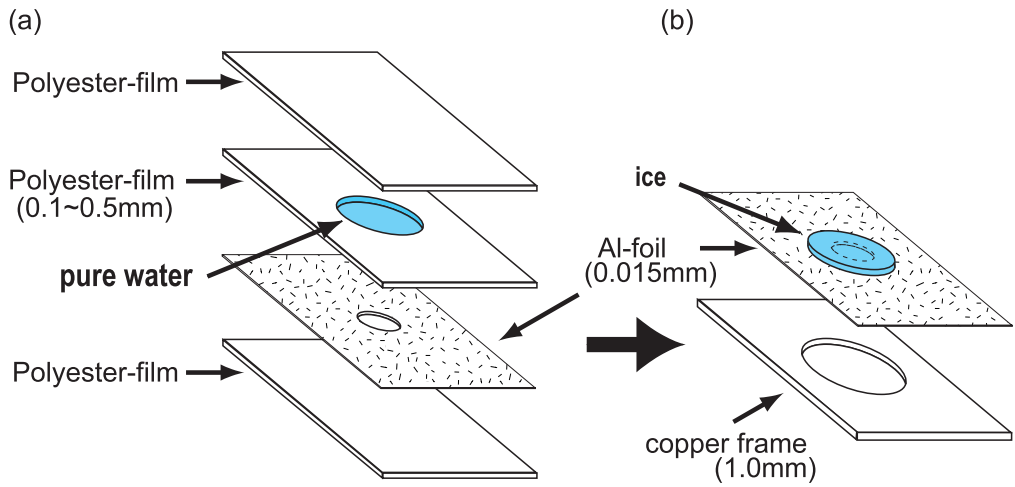,height=6.4cm}
\end{center}
\caption{
The process to make an ice target is shown.
(a) Pure water, 15 ${\rm \mu m}$-aluminum foil with a hole (10mm
in diameter) and three polyester films are stacked and frozen in a
refrigerator. (b) Then, the aluminum foil with ice is mounted on a 1
mm-thick copper frame after all polyester films are removed.
}
\label{fig:makeice}
\end{figure}

The process to make an ice sheet-target is illustrated in
Fig.~\ref{fig:makeice}.  
At first, 15 ${\rm \mu m}$-thick aluminum foil with a hole of 10
mm$^\phi$ is prepared and a spacer polyester film with a larger hole
is placed on top of it. Since the target thickness is mainly
determined by the thickness of the spacer, an appropriate thickness
should be selected. The polyester films  with thicknesses $0.1\sim0.5$
mm were used to make $10\sim50$ mg/cm$^2$ thick targets.  
They are further covered by polyester films from both sides 
as shown in Fig.~\ref{fig:makeice}(a).  

As a next step, a few drops of pure water are carefully poured into the 
hole space made by the spacer. The amount of water is adjusted
depending on the desired thickness of the target taking the density of
the ice (0.92 g/cm$^3$ at 0$^{\circ}$C) into account.  
The whole stack is cooled in the freezer box of a refrigerator. 
Then, keeping the ice frozen, the covering polyester films and the 
spacer polyester film are carefully removed.  
The minimum thickness of 10 mg/cm$^2$ has been achieved with a skilled 
and careful treatment. The aluminum foil with the ice sheet on top of
it is mounted on a pre-cooled, 1mm-thick copper frame with a 20
mm$^\phi$ hole as shown in Fig.~\ref{fig:makeice}(b). 

Finally, the mounted ice sheet is slowly cooled by nitrogen vapor from 
liquid nitrogen (LN$_{2}$) in order to avoid cracking. Cracks in the
ice sheet cause the deterioration of thermal conductance. 
The direct contact between an ice sheet and the copper
frame should be avoided since thermal stress due to the difference of
thermal compressibilities between ice and copper causes cracking
of the ice. The soft aluminum foil plays a crucial role 
in reducing the stress and keeping enough thermal conductivity.
It is better to reduce the amounts of material around the ice sheet
in order to avoid the background events from the beam halo.
The thin aluminum foil is also preferable from this view point.  
The hole size of 10 mm$^\phi$ is much larger than the beam spot size
of 1 mm$^{\phi}$.

\section{Target Cooling System}

\begin{figure}
\begin{center}
\epsfig{file=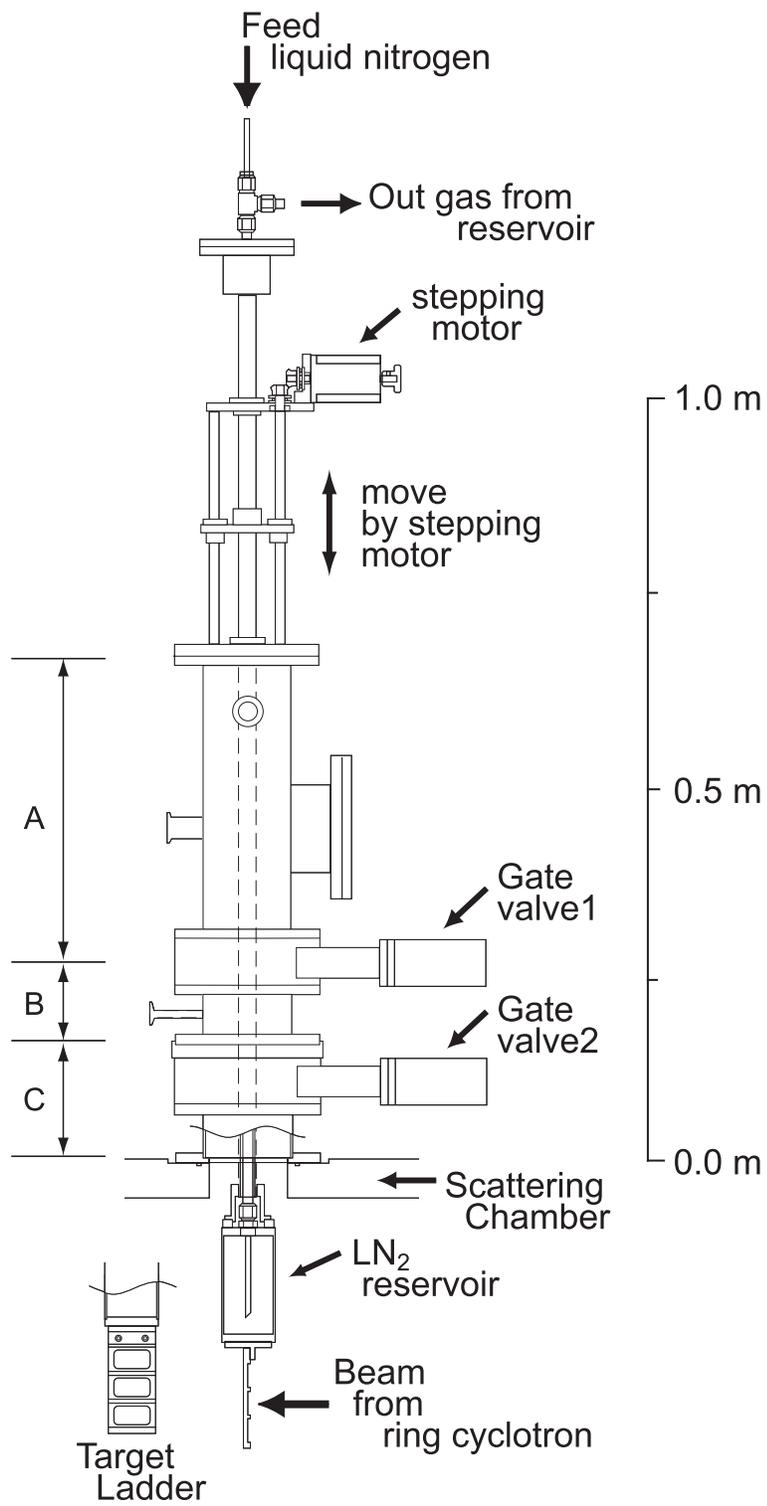,height=20.cm}
\end{center}
\caption{
Schematic view of the target cooling system. It is installed on the top of
the scattering chamber.
}
\label{fig:appra}
\end{figure}

A schematic view of the cooling system for ice targets is shown in
Fig.~\ref{fig:appra}.
LN$_{2}$ is fed from the top of the apparatus and stored in
the reservoir above the target ladder made of copper. 
LN$_{2}$ is introduced to
the reservoir through the central part of the triple concentric pipe, 
and vaporized nitrogen gas is exhausted through the middle pipe. The
space between middle and outside pipes is kept in vacuum to insulate
the heat.  Since the target ladder is directly connected to the
reservoir, it is cooled down to about 90 K, a temperature only
slightly higher than that of LN$_{2}$ (77 K).
The temperature is monitored by thermocouples. An automatic liquid
nitrogen feeding system is used. Liquid nitrogen is fed in every 3
hours and the feeding stops when the reservoir becomes full. 
The consumption of liquid nitrogen is 1.5 kg per 3 hours.
This system is 
mounted on top of the scattering 
chamber of the spectrometer Grand Raiden \cite{mamo}. 
The vertical position of the target can be changed by using a
stepping motor, which is controlled remotely via the RS-232C
connection. 

The installation of the ice target is done in the following procedure.  
Initially the target ladder is pre-cooled by LN$_{2}$  without any ice
targets, and the parts denoted as A and B in Fig.~\ref{fig:appra} are 
removed from the scattering chamber in order to mount the ice targets. 
This process is done without breaking vacuum of the scattering chamber 
or the target container (part A) by closing the gate valve 1 and 2; 
only the vacuum of the part B is broken. 
After nitrogen gas is introduced to the part A, 
the target ladder is pushed out of the part A through the gate valve
1. Then the ice target cooled by the nitrogen vapor from LN$_{2}$ is 
quickly mounted on the target ladder.
As soon as the target is mounted, the target ladder is pulled back 
and the part A is pumped out to lower than $10^{-3}$ Pa.
Finally, the parts A and B are connected with the part C on the
scattering chamber and the target ladder is installed into the
scattering chamber.

\section{Experiment and Results}

The $^{16}$O$(p,p')$ spectra were measured using
a 392~MeV proton beam at  RCNP ring-cyclotron facility.
A proton beam extracted from the ECR (Electron Cyclotron Resonance)
ion source was accelerated by the $K$~=~120~MeV AVF (Azimuthally Varying 
Field) cyclotron, and was further boosted  to higher energies
by the $K$~=~400~MeV ring-cyclotron. The emmitance of the 
beam was defined by slits between the AVF cyclotron and the
ring-cyclotron. The extracted beam was achromatically 
transported from the ring-cyclotron to the scattering
chamber of the two-arm magnetic spectrometer system (for details, 
see, for example, Ref.~\cite{Noro}). 
The ice target was bombarded by
a proton beam with a maximum intensity of 10 nA. The beam spot size 
was about 1 mm in diameter.
The vacuum in the scattering chamber was kept lower than
$2\times10^{-3}$ Pa. 
Scattered protons  from $^{16}$O were momentum analyzed by the
high-resolution spectrometer Grand Raiden \cite{mamo} 
placed at several angles between 2.5$^{\circ}$ and 14.0$^{\circ}$.  
Protons scattered by $^{1}$H were measured by 
using a large acceptance spectrometer (LAS) \cite{matsuoka}.  
A proton beam was stopped by a Faraday cup in the scattering chamber
for the measurements at backward angles from 4.0$^{\circ}$ to
14.0$^\circ$. In the measurements at forward angles from 2.5$^\circ$ 
to 4.0$^\circ$, another Faraday cup placed between Q1 and SX magnets
of Grand Raiden was used in order to avoid interference with scattered
protons.

\begin{figure}
\begin{center}
\epsfig{file=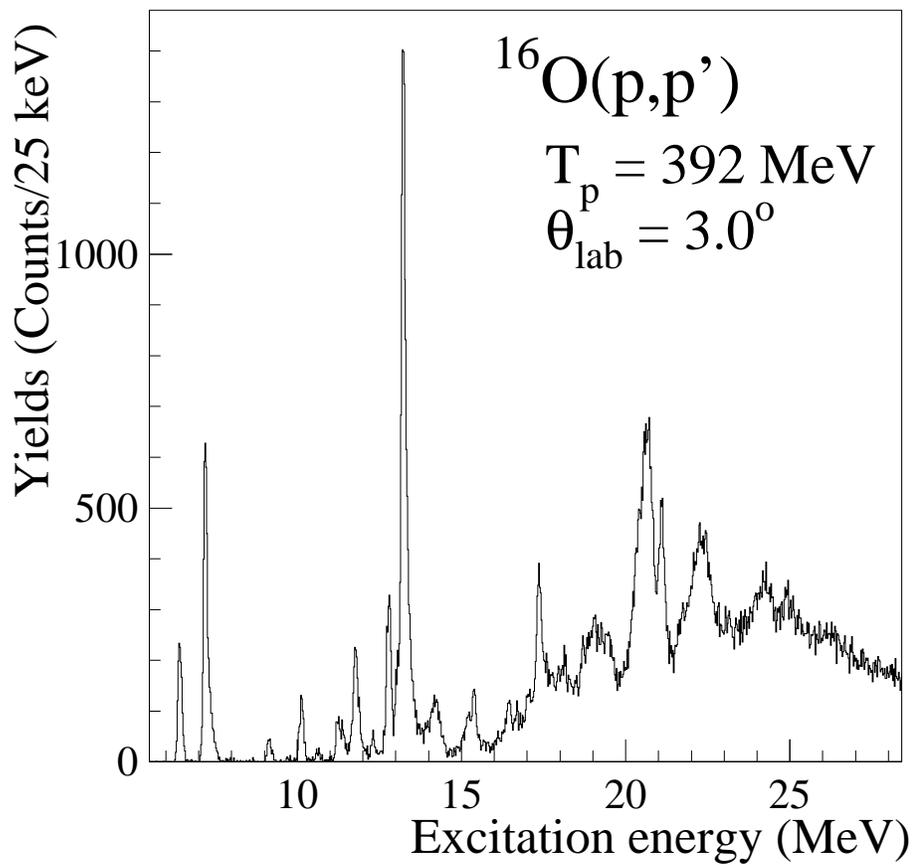,height=12.5cm}
\end{center}
\caption{Typical spectrum of inelastic proton scattering at the
bombarding energy of 392 MeV. An ice target with a thickness of 29.7
mg/cm$^2$ is used. The energy resolution obtained is 150 keV (FWHM).}
\label{fig:icespect}
\end{figure}

A typical spectrum of the inelastically scattered protons 
from $^{16}$O is shown in Fig.~\ref{fig:icespect}. The energy
resolution was 150 keV (FWHM). The spectrum consists of discrete
levels at lower excitation
energies and broad resonance bumps at higher energies.  
Since the target was kept near to the LN$_{2}$ temperature, 
contaminations due to the congelation of residual gases on the target
surface may happen in principle. However, we could not recognize
any significant peaks due to the contaminations, suggesting that the
spectrum is background-free. 

\begin{figure}
\begin{center}
\epsfig{file=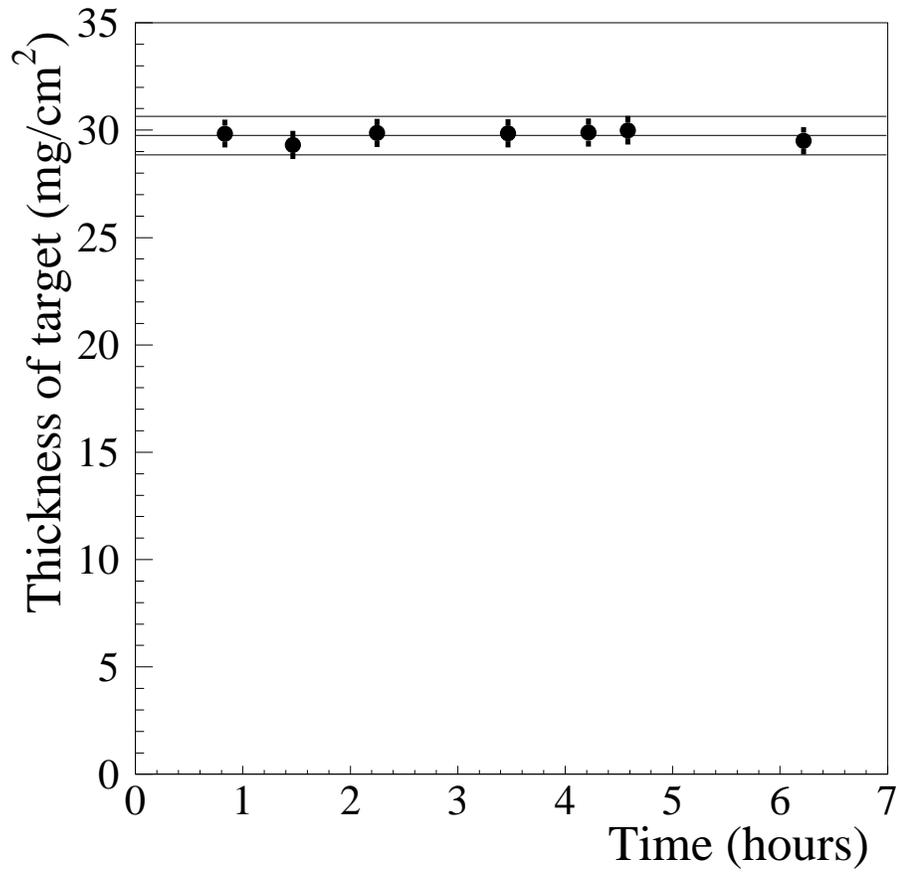,height=12.5cm}
\end{center}
\caption{
Stability of the ice
target thickness during irradiation by a proton beam
with intensity up to 10 nA.  The horizontal axis is total irradiation
time.  The vertical axis is measured target thickness.
Three lines, shown to guide the eye, indicate the mean thickness and 
$\pm 3\%$.}
\label{fig:thick}
\end{figure}

The target thickness can be changed by both the congelation and the
sublimation process. In order to check the stability of the target
thickness,  elastic scattering events from hydrogen were
monitored by LAS~\cite{matsuoka} placed at
$\vartheta_{lab} = 59.5^\circ$ for 7 hours in the experiment.
Due to the large momentum transfer in the backward proton-proton
scattering, the proton-proton elastic peak was unambiguously
identified. Thus, we could easily extract the yield of
the elastic peak on the continuum 
by assuming the smooth distribution of the quasi-elastic events.
The absolute cross section of proton-proton
elastic scattering at $\vartheta_{lab} = 59.5^\circ$ at 392 MeV 
was calculated using the SAID program \cite{arndt} based on the 
partial-wave analysis of various experiments. 
By comparing the calculated cross section with
the measured result from LAS, the thickness of the ice target 
was determined to be 29.7 mg/cm$^2$.  
The variation of the target thickness is shown in
Fig.~\ref{fig:thick} as a function of the irradiation time. 
The target thickness remained stable (within the 2.5 \% 
measurement uncertainty) during the entire 7 hour irradiation time.

\section{Summary}

A method to prepare a windowless, self-supporting and low-cost ice 
target has been described. The produced ice target was successfully used
to measure the spectra of inelastic proton scattering from $^{16}$O.
A ice target of about 30 mg/cm$^{2}$ worked stably.  
The target thickness was monitored for 7 hours by measuring p+p
elastic scattering and no meaningful variation of the thickness was
observed.  
Furthermore, no obvious effect of contaminations due to the
congelation of residual gases on the target was detected. 
Since a thin ice target was made, 
measurements of charged decay particles in coincidence with
inelastically scattered particles are possible.

\begin{ack}
The authors would like to thank Prof. Y.~Fujiyoshi, Dr. K.~Mayanagi,
and the Morishima-Seisakuzyo, Co., Ltd. for valuable discussions during the
development of the presented target system. 
We gratefully acknowledge the RCNP cyclotron staff for their support. 
This research program was supported in part by the Research
Fellowships of the Japan Society for the Promotion of Science (JSPS) for
Young Scientists. 
\end{ack}



\clearpage

\begin{table}
\caption{Existing oxygen targets and their characteristics.}
\label{table:targets}
\vspace{5pt}
\begin{center}
\begin{tabular}{llll}
\hline
Type& Merit& Demerit& Ref.\\
\hline
\vspace*{5pt}
\begin{minipage}[t]{2.8cm}
Metal oxide\\
(BeO, Li$_2$O, ...)
\end{minipage}&
Easy to obtain.&
\begin{minipage}[t]{4.5cm}
Large background \\
\hspace*{17pt}events from metal base.
\end{minipage}&
\cite{seifert}
\\
\hline
\vspace*{5pt}
\begin{minipage}[t]{2.8cm}
Gas target\\
{\scriptsize (Max $\sim10\,{\rm mg/cm^2}$)}
\end{minipage}&
Relatively clean.&
\begin{minipage}[t]{4.5cm}
Difficult to obtain\\
\hspace*{30pt} enough thickness.\\
Background events\\
\hspace*{30pt} from window foil.
\end{minipage}&
\cite{djalali}\\
\hline
\vspace*{5pt}
\begin{minipage}[t]{2.8cm}
Gas-jet target\\
Differential\\
\hspace*{5pt}pumped target\\
{\scriptsize (Max $\sim1\,{\rm mg/cm^2}$)}
\end{minipage}&
\begin{minipage}[t]{3.1cm}
Clean.\\
No window foil.
\end{minipage}&
\begin{minipage}[t]{4.5cm}
Expensive,\\
Limited areal density.
\end{minipage}
&
\begin{minipage}[t]{1.3cm}
\cite{gasjet1,gasjet3,gasjet6}\\
\cite{diffpump}
\end{minipage}\\
\hline
\vspace*{5pt}
\begin{minipage}[t]{3.0cm}
Water fall\\
\hspace*{10pt}
{\scriptsize ($25\sim140\,{\rm mg/cm^2}$)}\\
\hspace*{3pt}Ice\\
\hspace*{10pt}
{\scriptsize ($\sim300\,{\rm mg/cm^2}$)}
\end{minipage}&
\begin{minipage}[t]{3.1cm}
Easy to subtract \\
\hspace*{8pt}background\\
\hspace*{16pt}from hydrogen.
\end{minipage}&
\begin{minipage}[t]{4.5cm}
Background events\\
\hspace*{30pt} from window foil.
\end{minipage}&
\begin{minipage}[t]{1.1cm}
\cite{water1,water2}\\
\\
\cite{ice1,ice2}
\end{minipage}\\
\hline
\end{tabular}
\end{center}
\end{table}

\clearpage

\clearpage

\clearpage



\clearpage

\end{document}